\documentclass[sigplan,10pt]{acmart}
\renewcommand\footnotetextcopyrightpermission[1]{} 

\setcopyright{none}

\usepackage{booktabs}   
\usepackage{subcaption} 

\usepackage{listings}
\usepackage{xcolor}
\definecolor{codegreen}{rgb}{0,0.6,0}
\definecolor{codegray}{rgb}{0.5,0.5,0.5}
\definecolor{codepurple}{rgb}{0.58,0,0.82}
\definecolor{backcolour}{rgb}{0.95,0.95,0.92}

\lstdefinestyle{mystyle}{
    backgroundcolor=\color{backcolour},   
    commentstyle=\color{codegreen},
    keywordstyle=\color{magenta},
    numberstyle=\tiny\color{codegray},
    stringstyle=\color{codepurple},
    basicstyle=\ttfamily\footnotesize,
    breakatwhitespace=false,         
    breaklines=true,                 
    captionpos=b,                    
    keepspaces=true,                 
    numbers=left,                    
    numbersep=5pt,                  
    showspaces=false,                
    showstringspaces=false,
    showtabs=false,                  
    tabsize=2
}

\usepackage{multirow}
\usepackage{tabularx,booktabs}
\usepackage{tikz}
\usepackage{enumerate}
\usepackage{natbib}
\usetikzlibrary{shapes.geometric, arrows}
\usetikzlibrary{positioning}
\tikzstyle{startstop} = [rectangle, text centered, draw=black, fill=white!30, minimum width=1.1cm, minimum height=0.8cm]
\tikzstyle{arrow} = [thick,->,>=stealth]
\usepackage{environ}
\usepackage{array}
\usepackage{caption}
\usepackage[inline]{enumitem}
\makeatletter
\newsavebox{\measure@tikzpicture}
\NewEnviron{scaletikzpicturetowidth}[1]{%
  \def\tikz@width{#1}%
  \begin{lrbox}{\measure@tikzpicture}%
  \BODY
  \end{lrbox}%
  \pgfmathparse{#1/\wd\measure@tikzpicture}%
  \BODY
}
\makeatother
\begin{document}

\title{
Performance Optimization using Multimodal Modeling and Heterogeneous GNN }


\author{Akash Dutta\textsuperscript{1}, Jordi Alcaraz\textsuperscript{2}, Ali TehraniJamsaz\textsuperscript{1}, Eduardo Cesar\textsuperscript{3}, Anna Sikora\textsuperscript{3} and Ali Jannesari\textsuperscript{1}}
\affiliation{
  \institution{\textsuperscript{1}Iowa State University, Ames, IA, USA}            
  \country{}                    
}
\email{{adutta, tehrani, jannesari}@iastate.edu}          

\affiliation{
  \institution{\textsuperscript{2}University of Oregon, Eugene, Oregon, USA}            
  \country{}                    
}
\email{jordia@uoregon.edu}          

\affiliation{
  \institution{\textsuperscript{3}Universitat Autònoma de Barcelona, Barcelona, Spain}            
  \country{}                    
}
\email{{Eduardo.Cesar, Anna.Sikora}@uab.cat}          


\begin{abstract}
Growing heterogeneity and configurability in HPC architectures has made auto-tuning applications and runtime parameters on these systems very complex.
Users are presented with a multitude of options to configure parameters. 
In addition to application specific solutions, a common approach is to use general purpose search strategies, which often might not identify the best configurations or their time to convergence is a significant barrier.
There is, thus, a need for a general purpose and efficient tuning approach that can be easily scaled and adapted to various tuning tasks. 
We propose a technique for tuning parallel code regions that is general enough to be adapted to multiple tasks.
In this paper, we analyze IR-based programming models to make task-specific performance optimizations.
To this end, we propose the \textit{M}ultimodal \textit{G}raph Neural Network and \textit{A}utoencoder  ({\tt MGA}) tuner, a multimodal deep learning based approach that adapts Heterogeneous Graph Neural Networks and Denoizing Autoencoders for modeling IR-based code representations that serve as separate modalities. 
This approach is used as part of our pipeline to model a syntax, semantics, and structure-aware IR-based code representation for tuning parallel code regions/kernels. 
We extensively experiment on {\tt OpenMP} and {\tt OpenCL} code regions/kernels obtained from PolyBench, Rodinia, STREAM, DataRaceBench, AMD SDK, NPB, NVIDIA SDK, Parboil, SHOC, and LULESH benchmarks. 
We apply our multimodal learning techniques to the tasks of 
\begin{enumerate*}[label=(\roman*)]
    \item optimizing the number of threads, scheduling policy and chunk size in {\tt OpenMP} loops and,
    \item identifying the best device for heterogeneous device mapping of {\tt OpenCL} kernels.
\end{enumerate*}
Our experiments show that this multimodal learning based approach outperforms the state-of-the-art in all experiments.
\end{abstract}

\keywords{Auto-tuning, Multimodal learning, Heterogeneous Graph Neural Networks, OpenMP, OpenCL}  

\maketitle
\pagestyle{plain}
\section{Introduction}
\begin{figure}
    \centering
    \includegraphics[width=0.45\textwidth]{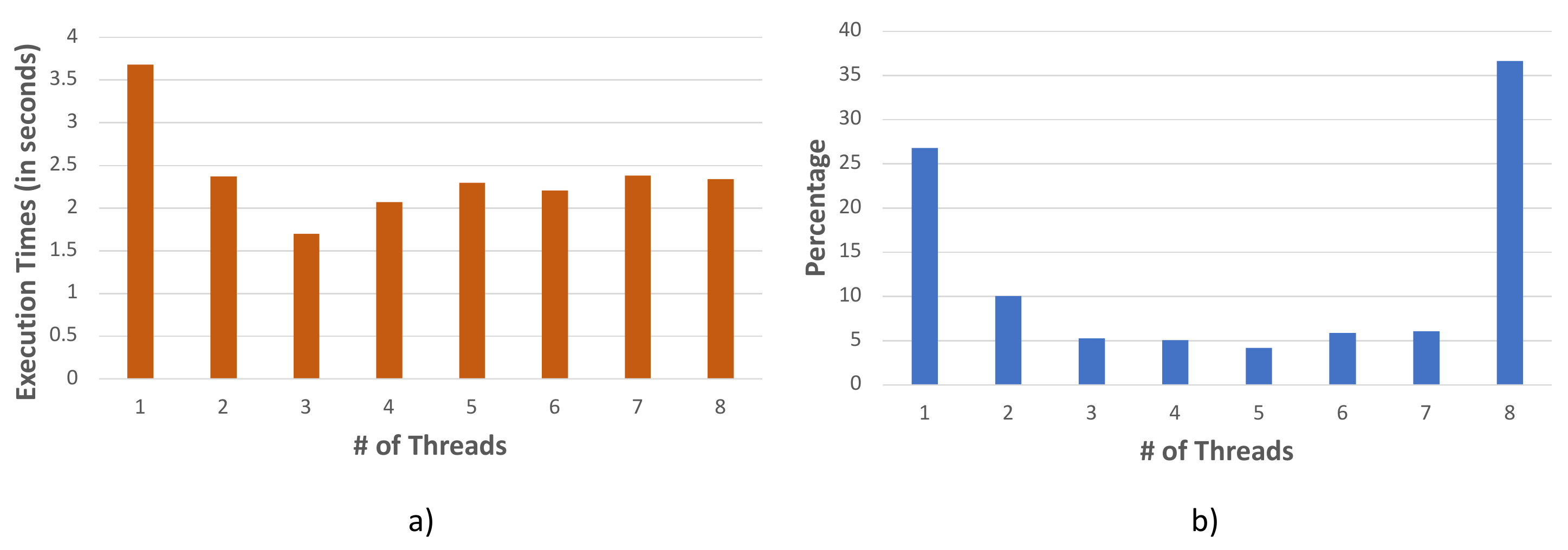}
    \caption{a) Execution times of {\tt kmeans} benchmark with different threads. b) Distribution of best thread counts across all {\tt OpenMP} loops and inputs in the dataset}
    \label{fig:motivation}
    \vspace{-5mm}
\end{figure}

With the onset of the exascale computing era, a lot of attention is now focused on HPC landscapes. 
However, the benefits of parallel programming is not just limited to supercomputers.
Most systems nowadays have multi/many-core architectures.
These hardware capabilities have led to the increased adoption of parallel programming models such as {\tt OpenMP}, and {\tt OpenCL} for writing parallel code.
Their shorter learning curves and ease of use has led to such programming models being used extensively not just for CPU programming, but also for programming accelerators such as GPUs.
Although these programming models have made it easier to convert serial code to parallel, they do provide users and programmers with various parameters that can be tweaked to highly impact performance.
However, selecting these parameters is often cumbersome and often needs expert guidance.
We aim to help address this by proposing a deep learning based IR-modeling technique for faster convergence and better results compared to state-of-the-art tools.
\newline
\textbf{\textit{Motivation.}} As an example, we evaluate the execution time of the {\tt OpenMP} version of the {\tt kmeans} kernel from the Rodinia \cite{che2009rodinia,che2010characterization} benchmark suite at different thread counts on an eight core system. 
We see significant difference in performance by varying the number of threads (Figure \ref{fig:motivation}a). 
There are four thread counts that achieve better performance than the default eight threads, improving execution time by upto 27\%. 
{\tt kmeans}, like many others, allows variable user inputs.
Repeating such a brute-force approach for a large set of applications and variable inputs is not feasible. 
The full extent of the tuning task at hand can be shown through Figure \ref{fig:motivation}b, where across 45 {\tt OpenMP} loops, and 30 different inputs, approximately 64\% combinations require tuning to identify the best thread count.
Larger search spaces would render such a brute-force approach unrealistic. 
Our process aims to ease this while achieving better results than existing auto-tuners.
\newline
\textbf{\textit{Prior Works.}} A majority of the performance optimization works follow either static or dynamic analysis based approaches.
Static analysis based works regularly target compiler optimizations which are a set of carefully handwritten heuristics \cite{venkatakeerthy2020ir2vec}.
Several works have used machine learning algorithms to improve the optimization decisions made by compilers \cite{stephenson2005predicting, grewe2013portable, kulkarni2013automatic, magni2014automatic, mendis2019compiler, haj2020neurovectorizer}.

Dynamic analysis based autotuners have to extensively execute source code and are historically search based tools \cite{ansel2014opentuner, tapus2002active, thiagarajan2018bootstrapping}. 
Recently several tools have employed Bayesian optimization based surrogate models for tuning purposes such as \cite{balaprakashytopt, menon2020auto, roy2021bliss}.
These approaches become quite expensive when inputs to code kernels vary with great regularity, and also suffer from the overhead of executing applications multiple times. 
A deep learning (DL)-based approach, on the other hand, can automatically prune a large search space by aggressively pruning non-beneficial points.

Several approaches using DL have been proposed for performance optimization tasks. 
Most of these propose a new method of representing code to achieve high-quality results \cite{brauckmann2020compiler, venkatakeerthy2020ir2vec, cummins2021programl}. 
These approaches, however, only considers task specific features inherent to one form of code representation. 
This exposes the shortcomings of each representation and leads to a loss of some syntactic, semantic, and structural characteristics of code.
Contrary to the trend of proposing a new code representation for such tasks, we propose combining more than one such representation, to use their individual strengths to overcome the shortcomings of the other.
To this end, we propose a code modeling technique that builds on existing representations and improves results by adapting multimodal learning to the task of code modeling.
\newline
\textbf{\textit{Our contributions.}} We propose the {\tt MGA} tuner, that models two dissimilar static code representations as separate characterizations of the same piece of code.
This allows the conjunctive modeling of multiple code representations targeted towards a common end goal. 
To this end, we propose modeling a distributed program vector and a graphical code embedding as different modalities of our multimodal learner. 
This can address the shortcomings of other unimodal approaches. 
In this learner, the code graphs will be modeled by a heterogeneous graph neural network, and the distributed vectors are modeled using a denoising autoencoder. 
Moreover, as static features themselves cannot model the execution behavior with multiple inputs, we augment these with performance counters (dynamic features).
Similar to some of the approaches discussed in the previous paragraph, our modeling technique is intermediate represenatation (IR)-based, making our code modeling language and architecture agnostic. 
We will show later that for tuning {\tt OpenMP} runtime parameters, our approach produces better results than state-of-the-art autotuners, while needing less executions.
We will also show that our approach outperforms existing techniques on {\tt OpenCL}-based heterogeneous device mapping tasks.
To summarize, our contributions are as follows:
\begin{itemize}
    \item Designing a new IR based hardware-independent multimodal code modeling technique that encapsulates syntactic, semantic, and structural code features.
    \item Developing heterogeneous graph neural network models for modeling flow graphs.
    \item Using denoising autoencoders for modeling distributed code vectors
    \item Designing a DL-based tuning approach for {\tt OpenMP} runtime parameters with geometric mean performance gains of $3.4\times$ while predicting {\tt OpenMP} threads and $2.23\times$ for predicting threads, schedule, and chunk size.
    \item Quantifying the impact of performance counters on DL-based performance tuning.
    \item Analyzing the $\mu$-architecture portability of our approach.
    \item Building a multimodal learner for the task of {\tt OpenCL} based heterogeneous device mapping achieving state-of-the-art accuracy of $\sim98\%$
\end{itemize}
\textbf{\textit{Outline.}} Section \ref{sec:background} outlines the topics of interest for this paper, followed by our approach and experiments in Sections \ref{sec:approach} and \ref{sec:experiments}, respectively. We outline related works in Section 5, and discuss and conclude our paper in Sections \ref{sec:discussion} and \ref{sec:conclusions}.
\section{Background and Overview}
\label{sec:background}
In this section, we briefly describe the ideas and concepts relevant to this work.
\subsection{Code Representations and Deep Learning }
Representation learning is being increasingly used for code modeling tasks. 
A lot of previous works have represented programs as a sequence of lexical tokens \cite{cummins2021programl}.
However, this fails to capture program structure. 
To overcome this, syntax as well as semantics based representations have been proposed \cite{allamanis2018survey, brauckmann2020compiler,raychev2015predicting,allamanis2017learning,dam2018deep}. 
But these methods do not take into account control, data, and call flows in a program. 
Several approaches have been suggested to represent these flows \cite{li2019graph,ben2018neural, steiner2021value}. 
However, these often lack the information provided by syntactic and structural modeling of source code. 

{\tt IR2Vec} \cite{venkatakeerthy2020ir2vec} is a flow-aware code representation that is not structurally aware. 
It is a scalable encoding infrastructure that represents programs as a distributed embedding in continuous space. 
{\tt PROGRAML} \cite{cummins2021programl} is another IR-based code representation that can model code flow information along with the code structure as multi-graphs. 
Each multi-graph has a vertex for instruction and control-flow edges between them. 
Data flow is represented by including separate vertices for variables and constants and associated data-flow edges to instructions. 
Call flow is represented by edges between callee functions and caller instruction vertices.

For modeling static code features, we represent code as a multimodal problem that considers these two embedding techniques as separate modalities allowing us to overcome the shortcomings of each modality on its own. 
This multimodal approach allows our model to learn the syntactic, semantic, and structural characteristics of code and extract relevant features from them.

\subsection{Performance Profiling}
Static analysis is a powerful method for analyzing program properties. 
But, dynamic analysis is often essential for understanding the execution behavior of programs with various input sizes. 
Performance profiling is a means to this end. 
It is widely used to analyze how code/code section impacts the hardware components.
Performance counters are used by developers to identify bottlenecks and scope of improvement in code execution.
In this work, we use such counters to study the impact of various inputs on code execution.
\textit{perf}, Likwid \cite{treibig2010likwid}, PAPI \cite{mucci1999papi} are a few commonly used tools for profiling. We use PAPI to profile each loop in Section \ref{sec:openmp_tuning}. 

\subsection{Graph Neural Networks}
\label{sec:background_gnn}
DL has revolutionized the application of machine learning in tasks that deal with data from Euclidean space. However, data is being increasingly generated from non-Euclidean space \cite{wu2020comprehensive}. Such data can more readily be represented as a graph.
Graph Neural Networks (GNNs) were proposed as a means of modeling such data.
Almost all GNNs are implemented using Message Passing Neural Networks\cite{gilmer2017neural} (MPNN).
The goal of these networks is to learn the latent space representation of each node through its neighbouring nodes in the graph. There are 3 main functions for constructing an MPNN: i) \textit{Message:} constructs communication between neighboring nodes, ii) \textit{Aggregate:} aggregates messages received from neighbouring nodes, iii) \textit{Update:} updates the target node embedding according to Message and Aggregate functions. 

Recent advances in GNNs have led to the proposal of \textbf{Heterogeneous Graph Neural Networks} \cite{wang2020survey}. 
Such models are used to accurately model diverse data with multiple relations.
Real world graphical data usually consists of different sets of entities and relations and cannot be effectively modelled by homogeneous GNNs due to differences in node and edge features, and dimensionality. 
To overcome this issue, heterogeneous GNNs were proposed.
In this work, we use heterogeneous GNNs to model our flow multi-graphs.
\subsection{Autoencoders}
\label{sec:dae}
Classic autoencoders are a type of deep learning model where the inputs and outputs are ideally the same.
Most autoencoders follow the classic \textit{encode-code-decode} setup.
Given an input, the input is passed through layers of fully connected neural networks (ANN/MLP), called the encoder, which aims to compress the input to a smaller dimension.
The encoder layers are followed by a code layer, which is usually a single MLP layer with a user-defined dimensionality (number of nodes).
The code layer are followed by the decoder layers, which are nothing but layers of MLPs.
The last layer of the decoder must have the same dimensionality as the input layer.
Autoencoders are usually unsupervised techniques, where these models usually learn to approximate the identity function.
Autoencoders are commonly used for feature selection and extraction.

\textbf{Denoising autoencoders (DAEs   )} provide a twist on classic autoencoders where the inputs are selectively corrupted by randomly modifying certain inputs.
The most common practice is to set a percentage of inputs to zero.
The target in this case becomes the uncorrupted inputs.
Because the training process for DAEs makes use of example/target pairs to gauge training quality, it becomes a self-supervised technique.
The main task for DAEs thus is compression.
In this work, we have used DAEs to model code vectors obtained through IR2Vec for feature extraction and compression.
\subsection{Multimodal Deep Learning}
\label{sec:mmdl}
Multimodal learning refers to relating information from multiple sources towards a common goal \cite{ngiam2011multimodal}. If there are multiple methods of modeling a target task, a problem can be assigned as multimodal, with each modeling technique defined as a unique modality.
 
Multimodal learning has thus far mostly been applied to audio and video analysis, speech synthesis, and gesture recognition tasks \cite{summaira2021recent}. 
For example, in image and video description tasks, the visual content and the associated textual description can be considered as different modalities with the same target -- to enable the viewer to perceive the content and meaning of the image/video. 
We take inspiration from these ideas and apply it to the task of code representation. 
A sequential and graphical code representation can represent different modalities of the same piece of code. 
The most common approach taken towards multimodal modeling is to obtain high level embeddings from different sources and associate them towards a common task. 
Generally, early fusion and late fusion are two techniques used for associating data from disparate sources in multimodal learning \cite{ramachandram2017deep}.
On a high level, early fusion can be thought of as feature level fusion, where data from multiple sources are integrated into a single feature vector, before being used as input to a machine-learning model.
Late or decision-level fusion refers to aggregating outputs from multiple models built on top of different modalities.
This is often used as errors from multiple models are usually unrelated and such a method is feature independent.
In this paper, we use late fusion for merging the outputs obtained by modeling two separate modalities.
Further details about multimodal learning and its applications can be found in this survey \cite{summaira2021recent} by Summaira et al.
\section{The MGA Tuner}
\label{sec:approach}
\begin{figure*}
    \centering
    \includegraphics[ width=0.85\textwidth]{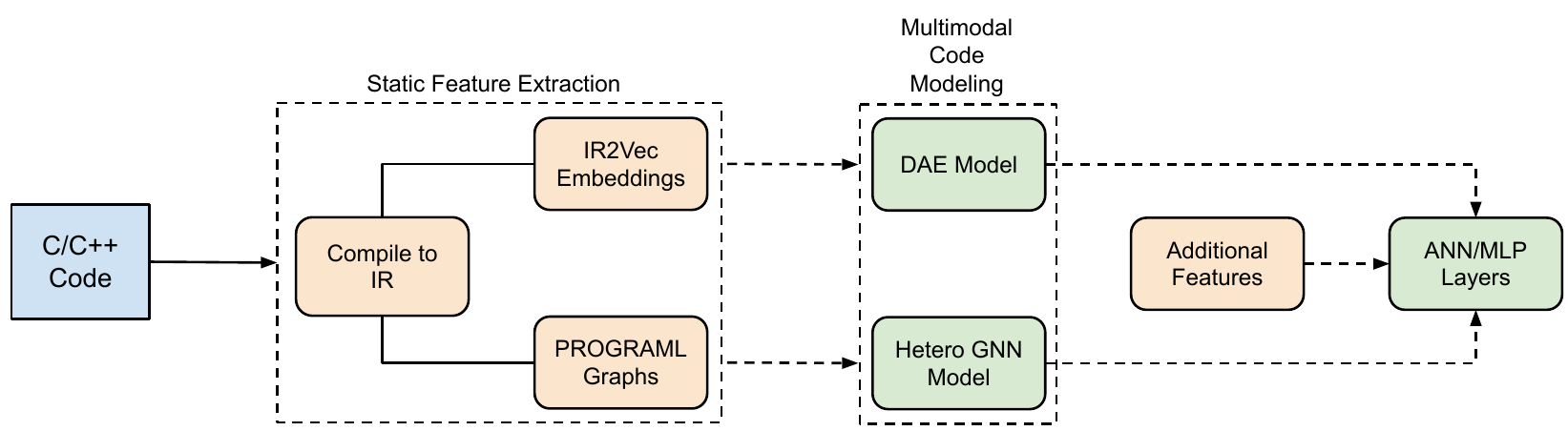}
    \caption{MGA Pipeline: An overview of the tasks in our Heterogeneous GNN based Multimodal DL tuner. The compiled IR is passed through IR2Vec and PROGRAML. The outputs are then passed through the DAE and heterogeneous GNN models respectively. Additional features are experiment specific as shown in Sections \ref{sec:openmp_tuning} and \ref{sec:opencl_tuning}.}
    \label{fig:hpdc_pipeline}
\end{figure*}
This section presents a novel framework that adapts advanced deep learning (DL) techniques to performance tuning tasks. 
We argue that for DL-based code modeling, code syntax, semantics, and structure are extremely important for proper understanding of code. 
However, using a combined representation would make modeling them too complex, lead to increased feature overlap, reduced specificity in identifying relevant features, and introduction of noise and conflicts.
To this end, we propose using two different code representations as separate modalities: i) a graphical code representation that can encode the code structure as a graph, ii) a distributed program vector representation that can encode syntactic and semantic features. 
A distributed vector representation can capture the relations within an instruction, but cannot effectively capture program structure. 
A graphical code representation, on the other hand, can fully capture code structure along with certain semantic features such as program flow. 
We aim to model the first modality using a heterogeneous graph neural network, and the second one using a denoizing auto-encoder (DAE).

However, these static code features are not sufficient for modeling the execution behavior of code with different inputs. 
Therefore, we augment these features with dynamic features such as performance counters to include additional information about program behavior/setup with varied inputs.
Figure \ref{fig:hpdc_pipeline} presents an overview of this tuning approach.

\subsection{Representing the Code}
\label{sec:approach_code_rep}
\begin{figure}
    \centering
    \includegraphics[scale=0.6]{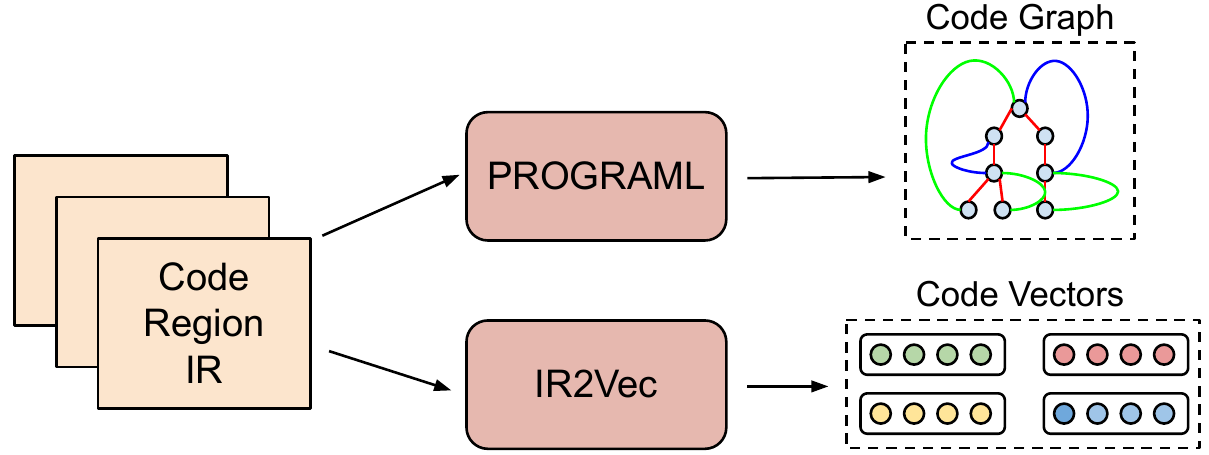}
    \caption{Multimodal code representation}
    \label{fig:mmdl_code_rep}
\end{figure}
Our multimodal code modeling is built on top of two very different state-of-the-art code representations ({\tt IR2Vec} and {\tt PROGRAML}).
Although the following part of the modeling process is done in parallel, we present these separately for improved readability and understanding.

To represent the first modality, each IR is passed through the {\tt PROGRAML} tool to obtain the corresponding code graphs, as shown in the upper half of Figure \ref{fig:mmdl_code_rep}. 
Along with representing the code structure, these code graphs also capture the data flow, control flow, and call flow in a single unified multi-graph.
To represent the second modality, the code region/loop IRs are used to generate a seed embedding vocabulary with the {\tt IR2Vec} encodings, as shown in the lower half of Figure \ref{fig:mmdl_code_rep}. 
This seed embedding is then used to obtain the code vectors used in the experiments. 
These code vectors and graphs are then passed through the code and performance modeling step as outlined in Section \ref{per_mod}.

\subsection{Performance Modeling} \label{per_mod}
The kernel IRs in our dataset are first transformed to a form usable by DL networks using techniques discussed in Section \ref{sec:approach_code_rep}.
Experiment specific features are also used to augment the static feature set.
In case of the {\tt OpenMP} experiments, performance counters are collected and used to incorporate the impact of various inputs on code execution.
For the experiments on {\tt OpenCL} kernels, we have used transfer and workgroup sizes as additional input features to our models.
These are discussed in further detail in Section \ref{sec:experiments}.
These, along with the static features, form the inputs to our models.
As shown in Figure \ref{fig:hpdc_pipeline}, our multimodal learning-based performance model can be abstracted into 3 high-level parts:

\textbf{\textit{Heterogeneous GNN modeling of flow graphs.}}
As mentioned in Section \ref{sec:background_gnn}, GNNs have been used for modeling graphical data. 
Heterogeneous GNNs have been successfully used for modelling real-world datasets with diverse node and edge attributes. 
The {\tt PROGRAML} flow graphs encapsulate the flow information in programs as three different types of relations. 
A homogeneous network cannot always fully incorporate multiple relationships in a multi-graph as shown in \cite{sun2019multi}. 
Therefore, to effectively model these flow multi-graphs, we have designed a heterogeneous GNN network capable of handling each of these three relations and the different types of nodes in the graph. 
This model is an agglomeration of three different GNNs to model each flow graph (data flow, control flow, and call flow). 
Each of these three sub-networks are homogeneous in nature as they are expected to model only a single relation and the associated nodes. 
Our heterogeneous GNN models these flow graph representations and their node features as shown in Figure \ref{fig:hpdc_pipeline}. 
Each homogeneous sub-network in the heterogeneous GNN network in this paper is a Gated Graph Convolutional Network \cite{li2015gated} with a "mean" aggregation scheme to group the node embeddings from each relation.

\textbf{\textit{Modeling code vectors using Denoising Autoencoders.}}
The dataset of code vectors obtained in Section \ref{sec:approach_code_rep} takes tabular structure, where each row in the table represents a sequence of code vectors.
The usual practice while working with tabular data is to use gradient boosted techniques such as XGBoost \cite{chen2015xgboost}, LightGBM \cite{ke2017lightgbm}.
Indeed such an approach has been used in \cite{venkatakeerthy2020ir2vec} for their modeling tasks.
However, due to the inherent difficulty of adapting GNNs and XGBoost as part of the same infrastructure, we have used denoising autoencoders as an alternative.
The best submission on a Kaggle competition with a tabular dataset \cite{kaggle_dae} highlighted DAEs as a possible alternative to gradient boosted algorithms.

This is a self-supervised technique, where we initially collect the code vectors in the training set and pass it through the DAE model.
Prior to modeling, the data is scaled into a standard normal distribution using Gaussian rank scaling.
During the encoding phase of the encoder-decoder architecture, we introduce "swap noise" into the dataset.
Imagine a table of data, where for any given column, a value in that column is replaced by a randomly sampled value from the same column, such that 10\% of values in a column has been modified.
This modified data is then input to the encoding phase of the DAE with the target of predicting the correct input.
This technique allows the DAE model to better learn the distribution of the dataset.

\textbf{\textit{Fully Connected Tuning.}} 
As discussed in Section \ref {sec:mmdl}, late fusion techniques were used to fuse the outputs from each modeled modality. 
The output tensors from the last level layers of the GNN and DAE networks are initially concatenated.
This fused tensor is then concatenated with additional features as mentioned before and detailed in Section \ref{sec:experiments}.
These additional features are performance counters for the {\tt OpenMP} experiments in Section \ref{sec:openmp_tuning}.
For the {\tt OpenCL} experiments, these are the transfer size and workgroup size.
Prior to concatenation, these features are normalized and scaled to a [0,1] range.
This feature vector is then fed as input to the fully connected  (dense/MLP) layers \cite{schmidhuber2015deep} as can be seen in Figure \ref{fig:hpdc_pipeline}.
These layers are trained with the target of identifying the best runtime configurations.
The fully connected MLP layers model all the aggregated features and classifies the loops/kernels and corresponding inputs to the appropriate configurations.
Our fully connected network consists of only one hidden layer.
We have consciously designed a small network  to reduce training time at source.
We show later that such a multimodal modeling technique produces much better results than other auto-tuners and state-of-the-art code representations using a single modality.
\section{Experiments}
\label{sec:experiments}
We validate our hypothesis on tasks using two programming models, {\tt OpenMP} and {\tt OpenCL}.
We chose to work primarily with {\tt OpenMP} as it is widely used in the parallel programming community, and can be easily compiled to their intermediate representations (IRs).
We additionally used {\tt OpenCL} to check the strength of our code representation and modeling technique.
Along with multiple benchmarks, we have worked with various input sizes to closely mimic real world scenarios.
We have compared our experiments with the state-of-the-art tools available in literature.
The setups for each experiment are detailed in the corresponding sections.

\textbf{\textit{Experimental Systems and Software.}}
The experiments in Section \ref{sec:th_pred} targets an 8-core Intel i7-10700K (Comet Lake) processor.
The experiments in Section \ref{sec:th_sch_chunk} target a $10$ core Intel Xeon Silver 4114 (Skylake) processor with two hyper-threads per core.
We work with a dataset generated on Intel Core i7-3820 CPU and AMD Tahiti 7970 and NVIDIA GTX 970 GPUs in Section \ref{sec:opencl_tuning}.
Code regions are compiled and extracted using Clang tools.
{\tt PyTorch} and {\tt Pytorch Geometric} libraries were used for building our DL models.

\textbf{\textit{Identifying and Selecting Benchmarks.}}
The first step in our pipeline centers around the appropriate selection of benchmarks for experimentation. 
The benchmark applications were selected to have sufficient variability amongst them.
We have used loops and kernels from multiple applications targeting domains ranging from arithmetic solvers to those targeting linear algebra, data mining, bioinformatics, fluid dynamics, image processing and others.
For the {\tt OpenMP} experiments, we used kernels from STREAM \cite{mccalpin1995stream}, DataRaceBench \cite{liao2017dataracebench}, Polybench \cite{pouchet2012polybench}, NAS \cite{barszcz1991parallel}, Rodinia \cite{che2009rodinia,che2010characterization}, and LULESH \cite{karlin2013lulesh, IPDPS13:LULESH} benchmarks. 
The {\tt OpenCL} experiments use kernels from the AMD SDK \cite{amd_sdk}, NPB \cite{seo2011performance}, NVIDIA SDK \cite{nvidia_sdk}, Parboil \cite{stratton2012parboil}, Polybench \cite{grauer2012auto}, Rodinia \cite{che2009rodinia} and SHOC \cite{danalis2010scalable} benchmark suites.
The benchmark applications used across all experiments are listed in Table \ref{tab:benchmark_list}.

\begin{table*}[ht]
\caption{List of benchmarks used in experiments}
\centering
\begin{tabular}{|p{0.16\linewidth}|p{0.8\linewidth}|}
\hline
\textbf{Benchmark Suite} & \textbf{Applications Selected} \\
\hline
Polybench \cite{yuki2015polybench} & 2mm, 3mm, atax, adi, bicg, cholesky, convolution-2d, convolution-3d, correlation, covariance, doitgen, durbin, fdtd-2d, fdtd-apml, gemm, gemver, gesummv, gramschmidt, jacobi-1d, jacobi-2d, lu, mvt, seidel-2d, symm, syrk, syr2k, trisolv, trmm\\
\hline
Rodinia \cite{che2009rodinia,che2010characterization} & b+tree, backprop, bfs, cfd, gaussian, hotspot, kmeans, lavaMD, leukocyte, lud, nn, nw, needle, particlefilter, pathfinder, srad, streamcluster\\
\hline 
NAS \cite{barszcz1991parallel} & BT, CG, EP, FT, LU, MG, SP\\
\hline 
STREAM \cite{McCalpin1995,McCalpin2007} & stream.c \\
\hline 
DataRaceBench \cite{liao2017dataracebench} & DRB045, DRB046, DRB061, DRB062, DRB093, DRB094, DRB121\\
\hline 
AMD SDK \cite{amd_sdk} & BinomialOption, BitonicSort, BlackScholes, FastWalshTransform, FloydWarshall, MatrixMultiplication, MatrixTranspose, PrefixSum, Reduction, ScanLargeArrays, SimpleConvolution, SobelFilter\\
\hline
NVIDIA SDK \cite{nvidia_sdk} & DotProduct, FDTD3D, MatVecMul, MatrixMul, MersenneTwister, VectorAdd\\
\hline
Parboil \cite{stratton2012parboil} & BFS, cutcp, lbm, sad, spmv, stencil\\
\hline
SHOC \cite{danalis2010scalable} & BFS, FFT, GEMM, MD, MD5, Reduction, S3D, Scan, Sort, Spmv, Stencil2D, Triad\\
\hline
LULESH \cite{karlin2013lulesh, IPDPS13:LULESH} & \\
\hline
\end{tabular}
\vspace{1mm}
\label{tab:benchmark_list}
\end{table*}

\subsection{OpenMP Tuning}
\label{sec:openmp_tuning}
In this section we have tried tuning {\tt OpenMP} runtime parameters for {\tt OpenMP} loops.
These parameters can highly impact performance on CPUs and we try to identify those configurations that lead to the fastest executions.
We initially compile the code to their IRs and model them as described in Section \ref{sec:approach}.
We augment these static code features with dynamic features in the form of performance counters.
Performance counters are necessary for this experiment to help analyze the impact of various inputs on an {\tt OpenMP} loop.

\subsubsection{Data Collection and Preprocessing} Initially, each application is instrumented to accept variable input at runtime. 
Additionally each {\tt OpenMP} loop is instrumented to call appropriate PAPI  \cite{mucci1999papi} APIs for profiling purposes.
Each instrumented application is then profiled for each input size and configuration.
This is a one time cost of creating the dataset and identifying the best configurations as labels of the dataset.
A major bottleneck of this process is the large number of available performance counters. 
All systems used for this experiment reports $>$50 preset counters.
We collected 20 PAPI counters based on the ideas presented in \cite{alcaraz2019hardware, alcaraz2021building, alcaraz2022predicting} for the Polybench suite. 
We extend and update these techniques to build our own dataset of {\tt OpenMP} loop signatures. 
For each loop, we used 30 input sizes ranging from $3.5$KB to $0.5$GB.
Profiling with multiple inputs provides insight into how these inputs impact the execution behavior of each {\tt OpenMP} loop.
The input sizes were selected with the intention of stressing each of the three cache levels (L1, L2, L3) to different degrees. 
This type of input-driven profiling lets us explore how varying runtime parameters can help alleviate latency issues.
However, including all counters while training our tuning model leads to a feature explosion and negatively impacts model convergence. 
To improve model convergence, we used Pearson's correlation \cite{benesty2009pearson} and identified five performance counters that are most correlated to execution time, and used these for training purposes. 
For the remaining applications, we only profile them to collect these five counters.
For an application with multiple {\tt OpenMP} loops, the associated counters and execution times are collected in a single run.
This implicitly accounts for the effect on hardware components a preceding loop might have on succeeding ones.
These steps reduce the profiling cost to a large degree.
The selected performance counters are L1, L2 cache misses, L3 load misses, number of retired branch instructions, and mispredicted branches across all loops, inputs and experiments.

\subsubsection{Setting up Baselines}
In this section, we have compared our results with three autotuners, {\tt ytopt}, {\tt OpenTuner}, and {\tt BLISS} and two state-of-the-art code representations {\tt PROGRAML} and {\tt IR2Vec}.
{\tt ytopt} \cite{balaprakashytopt} and {\tt BLISS} \cite{roy2021bliss} are autotuners based on Bayesian optimization.
{\tt OpenTuner} \cite{ansel2014opentuner} is a search-based autotuner which employs various search techniques such as AUC Bandit, Nelder-Mead, Torczon hillclimbers, etc.
{\tt ytopt} and {\tt OpenTuner} have been previously used for a variety of tuning tasks \cite{koo2021customized, wu2021autotuning, zhang2018graphit, han2020unicorn, hagedorn2018high} and were hence chosen as baselines for this paper.
{\tt BLISS} represents a more recent state-of-the-art autotuner based on Bayesian optimization.
We have also compared against unimodal DL approaches that uses only {\tt PROGRAML} \cite{cummins2021programl} or {\tt IR2Vec} \cite{venkatakeerthy2020ir2vec} as the code representations of choice.
\subsubsection{OpenMP Thread Prediction}
\label{sec:th_pred}
One of the most widely used techniques for improving the performance of {\tt OpenMP} code is by varying the thread-level parallelism.
Simply allocating more threads to a workload might not produce the best results as shown in Figure \ref{fig:motivation}.

The heterogeneous GNN model used across our experiments consists of three homogeneous GNN models.
We experimented with a few popular graph neural networks: graph convolution networks (GCNs) \cite{kipf2016semi}, graph attention networks \cite{velivckovic2017graph}, GraphSAGE \cite{hamilton2017inductive}, and gated graph neural networks (GGNN) \cite{li2015gated}. 
We observed that using GGNNs for modeling each relation in the flow graphs produces the best end results. 
The {\tt IR2Vec} embeddings are modeled using DAE layers with Sigmoid activation function as described in Section \ref{per_mod}. 
The output tensors from these models are then fused and concatenated with the performance counters and fed into the MLP layers to predict the number of threads.
This model is optimized with the AdamW optimizer \cite{loshchilov2018fixing}. 

To evaluate our model performance, we perform 5-fold cross validation. 
\begin{figure}
    \centering
    \includegraphics[width=0.47\textwidth]{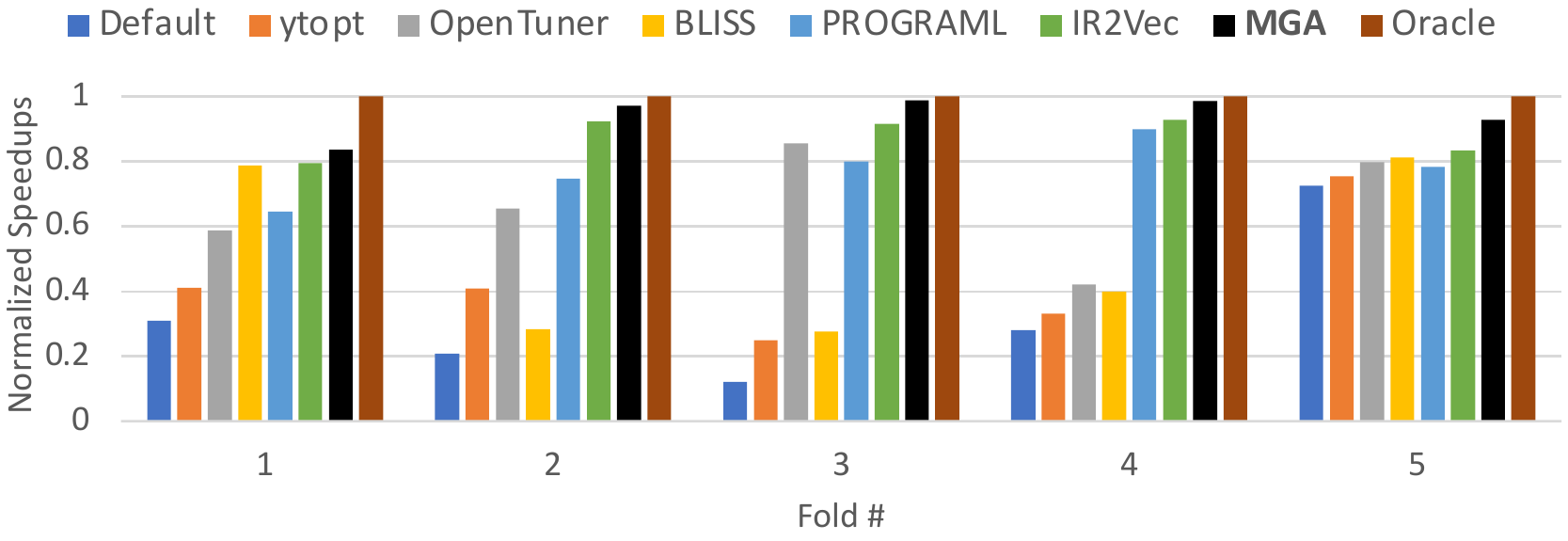}
    \caption{Thread Prediction: Normalized speedups (with respect to oracle speedups) per validation fold. The MGA tuner produces speedups of $2.71\times$, $4.68\times$, $8.09\times$, $3.51\times$, $1.31\times$ for each fold over default execution with all threads. Default speedup is always $1.0\times$.[Higher is better]}
    \label{fig:thread_pred_speedups}
\end{figure}
Here, we run the same experiment five times, where the five validation folds are mutually exclusive sets and the union of these five sets equal the set of all loops in the dataset. 
For each validation fold, the other loops in the dataset (four-fifth of all loops) are assigned to the training set.
The model is then iteratively trained and validated five times to ensure coverage of all loops in the dataset.
In the absence of a designated, representative test set, k-fold cross validation allows us to test the skill of the model on unseen data.
Our model achieves geometric mean accuracy of 86\% in identifying the best threads across five folds.

The results in Figure \ref{fig:thread_pred_speedups} show that our approach performs better compared to other approaches.
For each loop in the validation set, we use the predicted configuration for each input to obtain the execution time for that combination, and calculate the speedup (speedup = $\frac{Runtime_{default}}{Runtime_{new}}$) with respect to the default execution time.
We repeat this process for each loop in the validation set.
The geometric mean of these speedups is presented as each bar for each fold in Figure \ref{fig:thread_pred_speedups}.

As mentioned before, we compared our approach with two unimodal approaches using {\tt PROGRAML} and {\tt IR2Vec} as the code representation of choice along with the tuners {\tt ytopt}, {\tt OpenTuner}, and {\tt BLISS}. 
We defined the same search space for these tuners.
These tools performed search space optimizations to predict better performing configurations.
We treated the tuners as black boxes, and simply provided the search space and target metrics.
In this experiment, speedups are calculated with respect to the execution time with default {\tt OpenMP} configurations (all threads, static scheduling, compiler calculated chunk size).
In three out of five folds, our approach produced normalized speedups of $\ge 0.95 \times$, and in one out of five folds normalized speedup between $0.9\times$ and $0.95\times$ of the oracle speedups.
The {\tt IR2Vec} tuner led to normalized speedups of $\ge 0.9 \times$, but $< 0.95 \times $ in three out of five folds, and had normalized speedups $< 0.85 \times$ in the remaining folds. The {\tt PROGRAML} tuner produced normalized speedups of $>0.85 \times$ in one out of five folds.
As seen in Figure \ref{fig:thread_pred_speedups}, {\tt ytopt} produced normalized speedups $> 0.75 \times$ in one out of five folds. 
{\tt OpenTuner} and {\tt BLISS} produced speedups of $>0.75 \times$ in two out of five folds. 
Our approach only shows reduced performance gains in one fold.
This is primarily due to the presence of the {\tt trisolv} kernel from Polybench. 
The serial version of {\tt trisolv} has better performance than the parallel version used in this paper. 
This worsens the result of fold one, as the DL model does not see similar trends for other loops based on code modeling and execution behavior. 

Amongst the considered tuning approaches, our method came closest to the oracle predictions.
Oracle predictions in this experiment are those configurations obtained by brute-force tuning.
{\tt ytopt}, {\tt OpenTuner}, {\tt BLISS}, the {\tt PROGRAML} tuner, the {\tt IR2Vec} tuner, and the {\tt MGA} tuner produced geometric mean speedups of $1.46\times$, $2.33\times$, $1.67\times$, $2.79\times$, $3.17\times$, and $3.4\times$ across all folds compared to oracle speedups of $3.62\times$.

\textbf{\textit{Importance of dynamic information.}}
\begin{figure}
    \centering
    \includegraphics[width=0.47\textwidth]{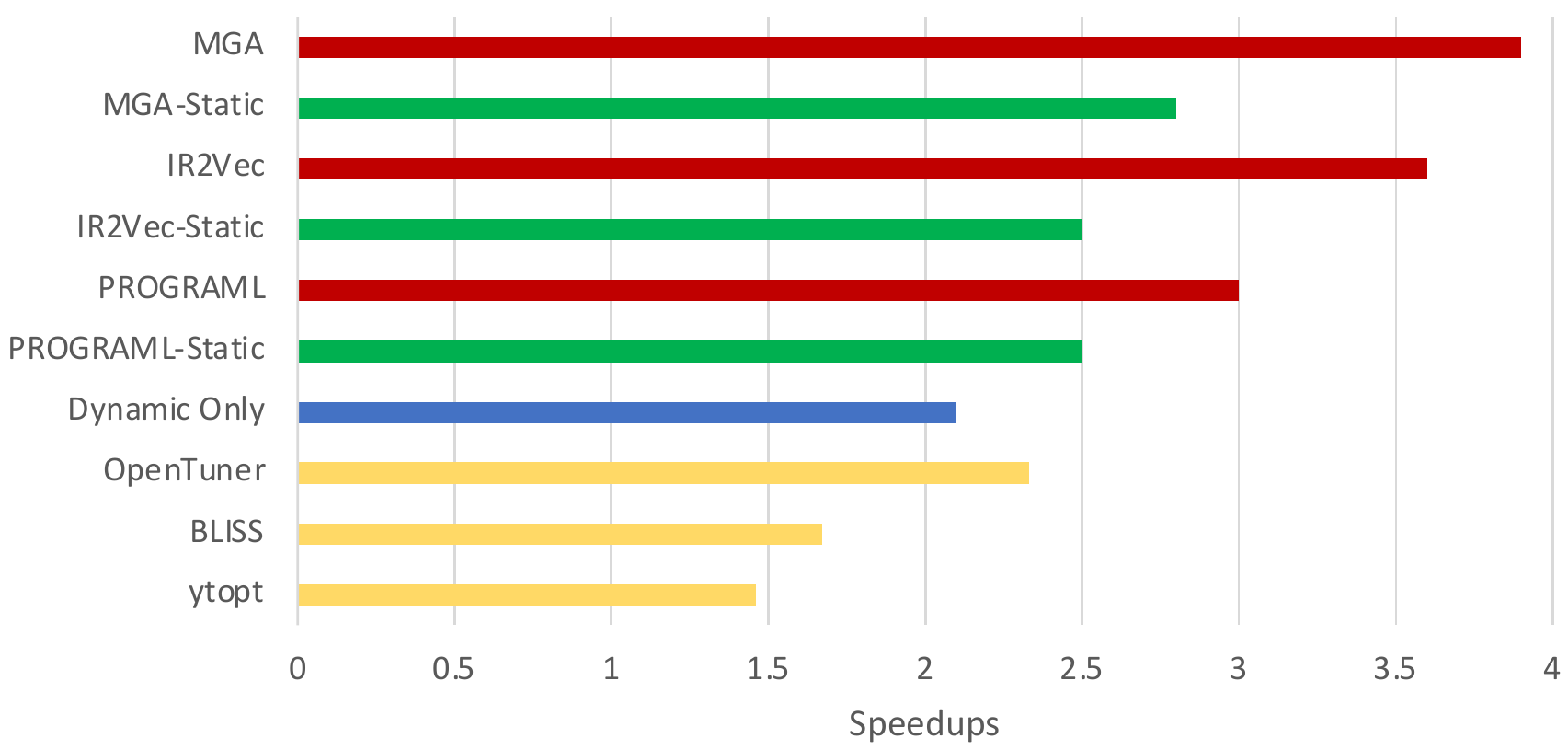}
    \caption{Thread Prediction: Impact of static and dynamic features. Red bars use both static and dynamic features. Green bars use only static features. The blue bar uses only dynamic features (perf. counters). Yellow bars are existing tuners in literature (added for comparison) [Higher is better].}
    \label{fig:perf_count_imp}
\end{figure}
Performance profiling is an overhead of our approach. 
However, we posit that modeling performance counters is essential for such DL-based tuning. 
We performed a set of ablation studies to validate this claim.
We trained three DL models with only static features and observed some performance degradation when performance counters were not a part of the feature set. 
The results from the validation set obtained by performing a randomized 80/20 split are shown in Figure \ref{fig:perf_count_imp}. 
Compared to achieved speedups of $3.9\times$, $3.6\times$, and $3.0\times$ by the {\tt MGA}, {\tt IR2Vec} and {\tt PROGRAML} models (uses both static and dynamic features) respectively, the speedups fell to $2.8\times$, $2.5\times$, and $2.5\times$ without performance counters.
This is expected as these static features do not explicitly provide information about the impact of varied inputs on execution.
In addition, we also train a model with only dynamic features. 
It showed the smallest speedups amongst all the DL-tuners designed in this paper, achieving speedups of only $2.1\times$. 
Therefore, static and dynamic features are both essential for tuning. 
We also compare these results with the {\tt ytopt}, {\tt OpenTuner}, and {\tt BLISS} predictions.
The tuners that use static-only features and both static and dynamic features have better performance than {\tt ytopt}, {\tt OpenTuner}, and {\tt BLISS}.

\begin{figure}
    \centering
    \includegraphics[width=0.47\textwidth]{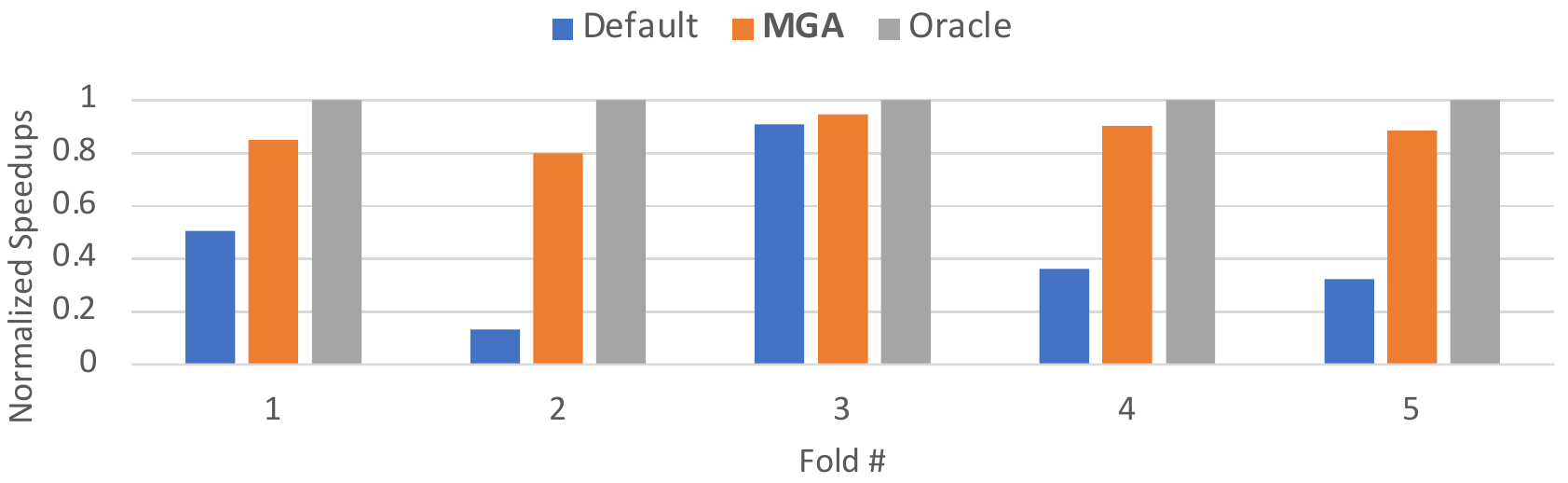}
    \caption{Thread Prediction on \textit{unseen loops and input size}. Speedups normalized with respect to oracle speedups. The MGA tuner produces $1.68\times$, $6.0\times$, $1.04\times$, $2.5\times$, $2.73\times$ speedups across five folds over default execution. [Higher is better]}
    \label{fig:threads_unseen_code_size}
\end{figure}

\textbf{\textit{Varying Input Sizes.}}
To evaluate the generalizability of our method, this section primarily evaluates how our model performs when \textit{both loops and input sizes are unknown}. 
We initially selected at random 20\% of the 30 input sizes considered in this paper, and set it aside for validation.
We then split the loops using 5-fold validation as described before.
Following this process, each validation fold now consists of unseen loops \textit{and} the unknown input sizes set aside before.
However, to reduce bias and preserve generalizability, the loops in these validation folds are different from the validation folds in the previous experiment (e.g. validation loops in fold one of this experiment is different from the validation loops in fold one of previous experiments).
In the previous experiments, only the {\tt OpenMP} loops were unknown in the validation set.
The model had been trained on the training set of {\tt OpenMP} loops and all input sizes.
In contrast, in this experiment, the model is trained on the {\tt OpenMP} loops in the training set and $80\%$ of the input sizes.
The loops in the validation sets and the unknown inputs are tested in this experiment and the results are shown in Figure \ref{fig:threads_unseen_code_size}. 
We observe that our model performs well producing geometric mean speedups of $2.35\times$ across all folds, compared to mean oracle speedups of $2.68\times$. 
There is some performance drop as input sizes highly impact performance counters and the best runtime configurations. 
Without prior knowledge of program behavior at these input sizes, the model performance suffers.

\subsubsection{Scaling up to a Larger Search Space}
\label{sec:th_sch_chunk}
\begin{figure*}
    \centering
    \includegraphics[width=0.9\textwidth]{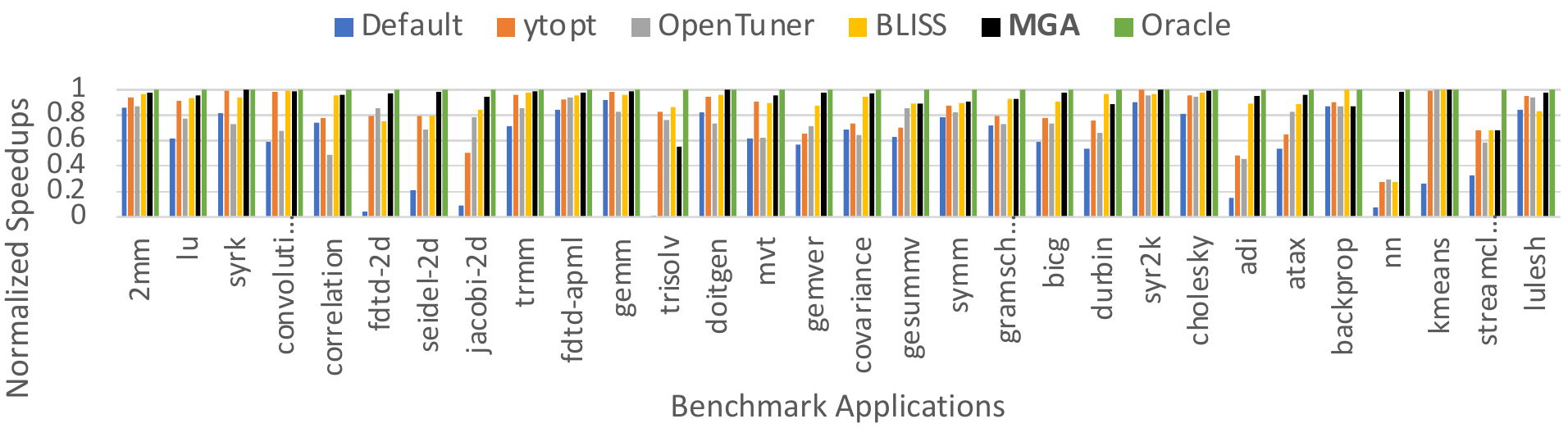}
    \caption{Normalized speedups (w.r.t. oracle) for each application for Section \ref{sec:th_sch_chunk} experiments. [Higher is better]} 
    \label{fig:thread_schedule_chunk_loov}
\end{figure*}
\begin{figure}
    \centering
    \includegraphics[width=0.48\textwidth]{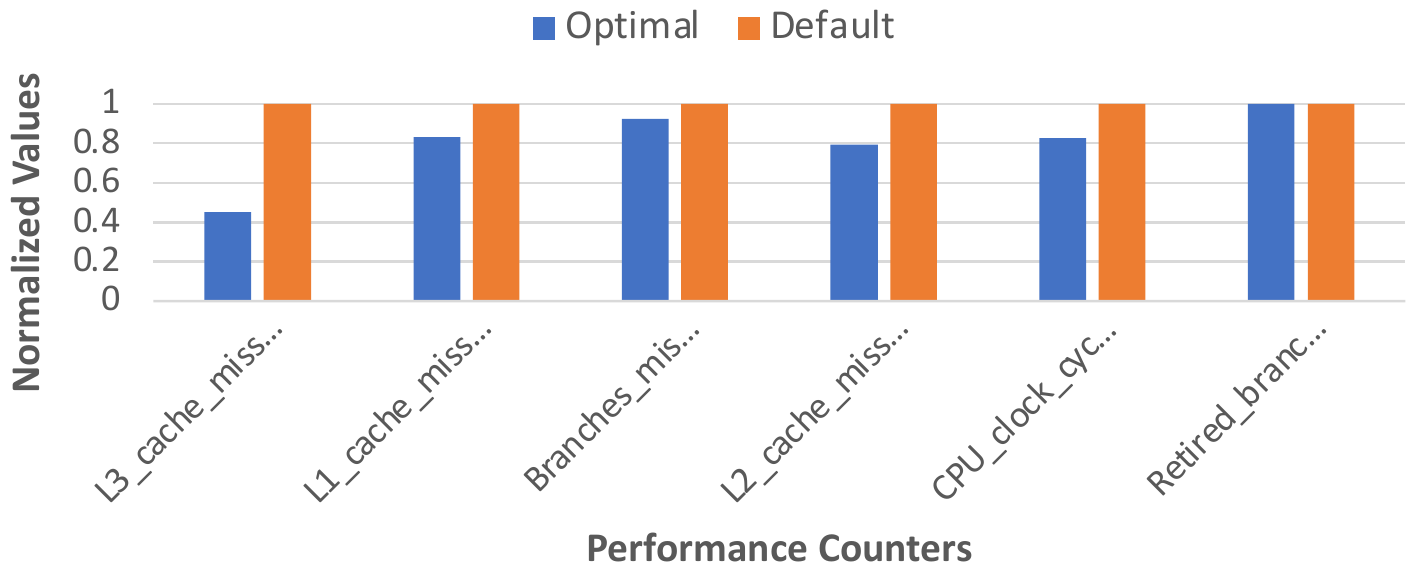}
    \caption{Normalized performance counter values for {\tt 2mm} benchmark: default (20 threads, static scheduling, dynamically calculated chunk sizes) vs predicted configuration (16 threads, dynamic scheduling, chunks of 8). [Lower is better]}
    \label{fig:perf_counter_def_vs_best}
\end{figure}

The experiments performed in previous sections have achieved good results. 
However, those search spaces are fairly small. 
In order to assess the scalability of our approach to larger search spaces, we experimented with tuning the number of threads, scheduling policy, and chunk sizes at the same time. 
The search space is defined in Table \ref{tab:search_space} using ideas from \cite{bari2016arcs, bari2019performance}. 
\begin{table}[ht]
\caption{Search Space for Experiment in Section \ref{sec:th_sch_chunk}}
\vspace{-1mm}
\centering
\begin{tabular}{|p{0.4\linewidth}|p{0.5\linewidth}|}
\hline
\textbf{{\tt OpenMP} Parameter} & \textbf{Parameter Values} \\
\hline
Threads & 1, 2, 4, 8, 12, 16, 20\\
\hline
Scheduling Policies & \textit{static}, \textit{dynamic}, \textit{guided}\\
\hline 
Chunk Sizes & 1, 8, 32, 64, 128, 256, 512\\
\hline 
\end{tabular}
\label{tab:search_space}
\end{table}
In this experiment, we have used a smaller subset of the applications considered in the paper and worked on benchmarks from Polybench and Rodinia.
We have additionally experimented with a proxy application, LULESH \cite{IPDPS13:LULESH, karlin2013lulesh} from the DARPA UHPC program.
To compensate for the reduction in the size of the dataset, we performed \textit{leave-one-out validation} instead of 5-fold validation. 
In this method of validation, we leave out data associated with one benchmark application (all loops in this application are present in the validation set) as the validation set and train our model on the rest. 
This process is repeated for each considered application. 
As shown in Figure \ref{fig:thread_schedule_chunk_loov}, this leads to normalized speedups of $>0.95\times$ of the oracle speedups in $21$ out of $30$ applications, and $>0.85\times$ normalized speedups in 28 out of 30 applications.
{\tt trisolv} is the worst performing application due to reasons discussed in Section \ref{sec:th_pred}.
Our approach outperforms {\tt ytopt}, {\tt OpenTuner}, and {\tt BLISS} in 28, 29, and 26 cases out of 30.
{\tt ytopt}, {\tt OpenTuner}, and {\tt BLISS} produce $>0.95\times$ of the oracle speedups in 7, 2, and 12 cases out of 30.
Overall, our model produces geometric mean speedups of 2.23$\times$ compared to oracle speedups of 2.38$\times$.
The improvement in performance of most kernels can be attributed to better cache performance, branch predictions, and load balancing. 
We show the impact of using the predicted {\tt OpenMP} configurations for this system on cache misses, clock cycles, and branch mispredictions compared to the default configuration of using all threads and static scheduling in Figure \ref{fig:perf_counter_def_vs_best} for the {\tt 2mm} kernel. 
There is a clear relation between improved performance and reduced cache misses, and branch mispredictions. In most kernels used in this paper, the profitable configurations lead to improvements in most of these factors, leading to improved performance.

\subsubsection{Analyzing $\mu$-architecture Portability}
\label{sec:perf_portability}
In this section, we analyzed if our auto-tuner can predict the number of threads on other $\mu$-architectures.
We re-used the model in Section \ref{sec:th_pred} (trained on data from Comet Lake $\mu$-architecture) to predict the number of threads on single-socket 8 core systems belonging to the Broadwell and SandyBridge $\mu$-architecture (access provided by Cloudlabs infrastructure \cite{Duplyakin+:ATC19}).
Limiting the scope of this experiment to testing hardware portability to single-socket 8 core systems helps us to directly use a pre-trained model without additional training (different core/socket count would necessitate re-training).
Static code graphs, sequential code vectors, and the performance counters from the target systems (Broadwell/Sand Bridge) were passed as inputs to the pre-trained model.
We validated this approach on 25 kernels from the PolyBench benchmark with STANDARD and LARGE inputs.
Similar to Section \ref{sec:th_sch_chunk}, we perform \textit{leave-one-out validation} for this experiment.
The data from the Comet Lake system was always used for training in this experiment.
For each validation fold, the validation kernel was executed twice on the target $\mu$-architecture to collect the necessary counters.
The L1, L2, and L3 cache counters were then computed as a function of the system cache sizes relative to the system on which the training data is collected (e.g. L1 cache misses for Sandy Bridge are computed as
$\frac{L1\_CM \times {L1\_cache\_size}_{SandyBridge}}{{L1\_cache\_size}_{CometLake}}$).
The branch misprediction counters were divided by the number of reference clock cycles.
These counters were then normalized to a [0,1] scale and fed into the model to predict the number of the threads.
This process completely removes the overhead of re-training models for other similar $\mu$-architectures.
\begin{figure*}
    \centering
    \includegraphics[width=0.9\textwidth]{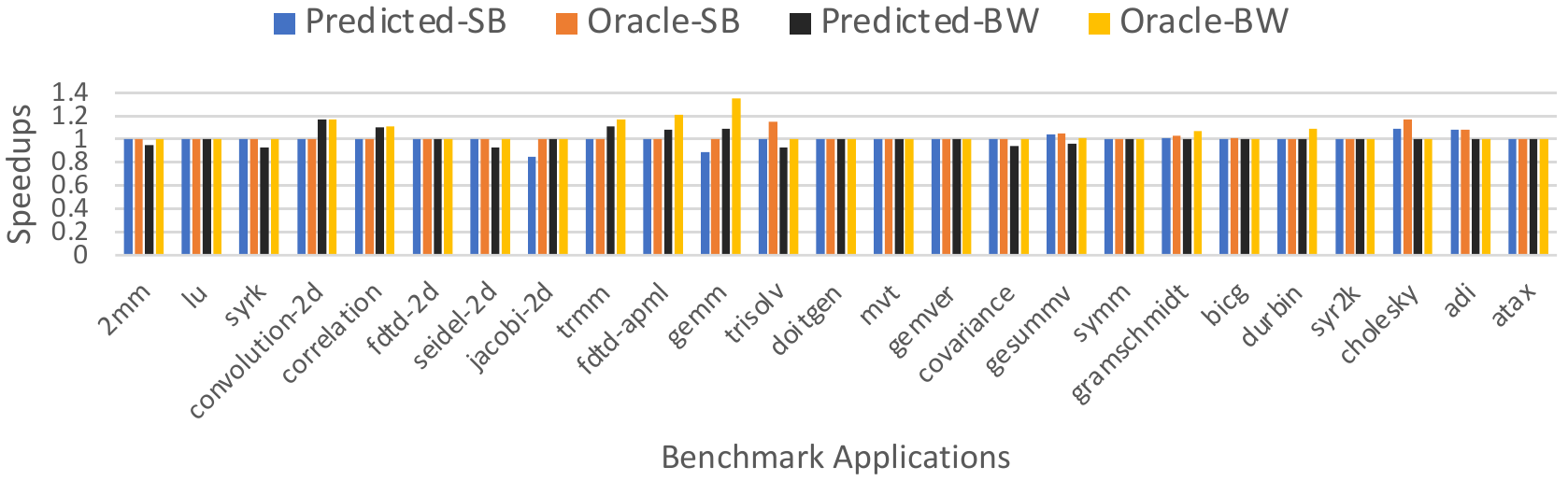}
    \caption{Speedups for experiment in Section \ref{sec:perf_portability}. Predictions for Broadwell (BW) and SandyBridge (SB) by model trained on data from Comet Lake [Higher is better]}
    \label{fig:broadwell}
\end{figure*}
As shown in Figure \ref{fig:broadwell}, our approach performs well while predicting the number of threads on Braodwell and Sandy Bridge with a model trained on data from Comet Lake.
In most cases, the predicted configurations lead to optimal/better performance.
We observed that only using static features leads to the model making similar predictions for Comet Lake, Broadwell, and Sandy Bridge.
Without modeling performance counters, the predictions for unseen loops are what it would be for the training $\mu$-architecture.
This led to degraded performance on the target $\mu$-architectures. \newline
\textbf{\textit{Observations and Analysis.}} For these experiments, our approach produces the best results overall.
The experiments considered in this section fall more under the scope of online/offline auto-tuning than compiler optimizations.
Search-based autotuners are more suited to this type of autotuning.
However, these do need to execute code multiple times to identify profitable configurations from the search space.
Our approach in this section needs only two runs (two runs are needed due to system limitations; the selected performance counters cannot be collected in one run; essentially, we only need one run when systems permit) during inference, irrespective of the size of the search space.
In comparison, {\tt ytopt}, {\tt OpenTuner}, and {\tt BLISS} require multiple runs to identify profitable configurations.
As an example, for the experiment in Section \ref{sec:th_sch_chunk}, we consider the time taken for tuning the {\tt 2mm} benchmark from PolyBench with LARGE input. 
The {\tt MGA} tuner takes approximately 90 seconds to identify near-optimum configurations during inference. 
This includes the time needed for profiling and prediction. 
{\tt OpenTuner} takes roughly 180 seconds (the specified time limit), {\tt ytopt} takes approximately 260 seconds to finish tuning with the maximum evaluations set to ten, and {\tt BLISS} takes around 220 seconds.
The settings for the three tuners were chosen to achieve a balance between acceptable results and speed of tuning.
Letting them run longer, although much more expensive, might improve results.
In conclusion, not only does our technique produce better results than the other approaches considered in this work, it is also faster than existing tuners.

\subsection{OpenCL Tuning}
\label{sec:opencl_tuning}
{\tt OpenCL} is another programming model that is widely used in parallel programming.
With this experiment we aim to validate if our approach works for a compiler optimization task for another IR-based programming model.
A commonly used task to validate the effectiveness of code representations and modeling is heterogeneous device mapping.
Grewe et al. \cite{grewe2013portable} proposed the device mapping task to map {\tt OpenCL} kernels to the CPU or GPU.
This task has been used in later works \cite{ben2018neural, cummins2021programl, venkatakeerthy2020ir2vec} to validate how good these are on this task.
We also use this task to benchmark our results against the state-of-the-art results obtained by PROGRAML and IR2Vec.

\subsubsection{Dataset}
We use the dataset published by Ben-Nun et al. \cite{ben2018neural} for this experiment. 
It has 256 unique {\tt OpenCL} kernels from seven benchmark suites comprising of AMD SDK, NPB, NVIDIA SDK, Parboil, Polybench, Rodinia and SHOC. 
The data size and workgroup size were varied for each kernel to obtain a labeled dataset with 670 CPU- or GPU-labeled data points for each of the two devices, AMD Tahiti 7970 and NVIDIA 970.
As this is a published dataset, no modifications were made to it, and performance counters have \textit{not} been used in this experiment.

\subsubsection{Heterogeneous Device Mapping}
For modeling purposes, we use similar techniques used in the previous section.
We initially use the extracted IR of the {\tt OpenCL} kernels in the dataset.
The IRs are then passed through {\tt PROGRAML} and {\tt IR2Vec} to obtain the code graphs and code vectors.
As before, the code graphs are modeled using Heterogeneous GNNs and the code vectors are modeled using DAEs.
For this experiment, the modeled outputs from the GNN and DAE models are concatenated as before.
In addition, we also add transfer and workgroup sizes from the dataset to the feature set before passing the feature set onto the fully connected MLP layers.
Following the techniques used in \cite{cummins2021programl, venkatakeerthy2020ir2vec}, we have also used ten-fold stratified cross-validation to evaluate our results.
We were able to replicate the experiments reported in \cite{venkatakeerthy2020ir2vec}, and these results are used to compare our results.
The results from \cite{cummins2021programl} are used directly to compare our results.
In this task, we validate if our approach can outperform the state-of-the-art to reinforce our hypothesis that existing code representations are good enough to be used in conjunction for better performance.

\begin{table}[ht]
\caption{Accuracy: Heterogeneous device mapping (CPU/GPU). All numbers are in percentage. Numbers in parenthesis are percentage improvements in accuracy of {\tt MGA} model over corresponding approaches.}
\centering
\begin{tabular}{|p{0.35\linewidth}|p{0.27\linewidth}|p{0.26\linewidth}|}
\hline
\textbf{State-of-the-art} & \textbf{NVIDIA GPU} & \textbf{AMD GPU} \\
\hline
Grewe et al. \cite{grewe2013portable} & 74.56 (31.3) & 70.29 (39.0)\\
\hline
DeepTune \cite{cummins2017end} & 80.88 (21.04) & 83.24 (17.37)\\
\hline
inst2Vec \cite{ben2018neural} & 82.65 (18.45) & 82.35 (18.64)\\
\hline
PROGRAML \cite{cummins2021programl} & 80 (22.38) & 86.6 (12.82)\\
\hline
IR2Vec \cite{venkatakeerthy2020ir2vec} & 89.68 (9.17) & 92.82 (5.26)\\
\hline 
\textbf{MGA} (ours) & \textbf{97.9} & \textbf{97.7}\\
\hline 
\end{tabular}
\label{tab:devmap_accuracy}
\end{table}

Our experimental setup leads to state-of-the-art results in identifying the correct device.
We achieve accuracy of 97.9\% and F1-score of 0.98 in identifying the best device on the NVIDIA GPU.
On the AMD GPU, we achieve accuracy and F1-score of 97.7\% and 0.97.
In comparison, PROGRAML achieves accuracies of 80\% and 86.6\% on the NVIDIA and AMD GPUs and corresponding F1-scores of 0.88 and 0.8.
IR2Vec (flow-aware representation) achieves accuracies of 89.68\% and 92.82\% on the NVIDIA and AMD GPUs.
Comparisons with other works on this dataset are shown in Table \ref{tab:devmap_accuracy}.
The accuracy numbers for Grewe at al. \cite{grewe2013portable}, DeepTune \cite{cummins2017end}, and inst2vec \cite{ben2018neural} are cited from \cite{venkatakeerthy2020ir2vec}.

We have also analyzed performance improvements due to the predictions by our model.
The speedups are calculated in comparison to static mappings as done in \cite{venkatakeerthy2020ir2vec}.
On the NVIDIA 970 system, our approach leads to speedups of $1.3\times$ compared to oracle speedups of $1.34\times$.
The oracle speedups are calculated by analyzing the execution time on the best device and comparing it to the static mapping baseline.
In comparison, the predictions in \cite{venkatakeerthy2020ir2vec} led to speedups of $1.26\times$.
On the AMD Tahiti system, our predictions lead to speedups of $1.62\times$ compared to speedups of $1.58\times$ produced by IR2Vec \cite{venkatakeerthy2020ir2vec} and oracle speedups of $1.66\times$. \newline
\textbf{\textit{Observations and Analysis.}} 
We have shown in this section that our approach produces better results than the state-of-the-art in this field without the need of a completely new code representation technique.
We analyzed our model's predictions to identify those cases where our model outperformed the state-of-the-art.
We were only able to replicate the experiments in \cite{venkatakeerthy2020ir2vec} (best results in existing literature) and our observations are with respect to this paper.
Our overall performance was better as our edge case predictions were better.
We noticed that our model outperformed in corner cases where kernels with small inputs were mapped to the GPU, and kernels with large inputs were mapped to the CPU.
As an illustrative example, consider the \textit{makea} kernel from the {\tt CG} benchmark in NPB.
In the dataset, this kernel gets mapped to a GPU with a small input class S, whereas the same kernel with much larger input class C gets mapped to the CPU.
This behavior can be due to the presence of multiple function calls from inside a loop.
The called functions, also have parallel loops in them.
This, we believe, creates an overhead which leads to faster execution on the CPU for larger inputs.
For the smaller inputs, the number of function calls are much less which does not create a bottleneck for GPU execution, leading to much faster execution on GPUs.
The MGA model is able to identify such edge cases as our approach can capture the characteristics of individual instructions and arguments along with the data, control, and call flows in a kernel.
\section{Related Work}
This paper proposes a new multimodal code representation technique built on top of state-of-the-art representation techniques and its usage for DL based tuning of runtime configurations for {\tt OpenMP} and {\tt OpenCL} kernels.
These programming models expose a number of configurations for runtime optimization. 
Thus auto-tuning is essential for identifying the optimum configurations. 

There already exists a large body of research on tuning runtime parameters or configurations for parallel code \cite{tapus2002active, mustafa2011performance, katarzynski2014towards, sreenivasan2019framework, gadioli2018margot, huda2016automatic}. 
{\tt OpenTuner} \cite{ansel2014opentuner} and {\tt ActiveHarmony} \cite{tapus2002active}  are autotuning frameworks for domain-specific tuning that is much faster than exhaustive search-based auto-tuners. 
These tuners employ a variety of search techniques for search space exploration and optimizations.

An alternative to search-based auto-tuning is to use machine learning based approaches.
Search-based auto-tuners mostly depend on manually or pre-defined heuristics to identify optimum points in the search space.
Such an approach iteratively explores the search space to identify patterns that might point to profitable configurations in the search space.
Such tuners, however, need to execute applications a number of times, which is often expensive.
ML tuners can reduce this exploration because of its pre-training and ability to associate similarities between applications. To this end, \cite{rameshka2019rigel, wang2014integrating} propose machine learning based approaches to {\tt OpenMP} autotuning. 
Artemis \cite{wood2021artemis} is an automatic parameter tuning framework that uses machine learning to predict the execution parameters of parallel regions.
{\tt ytopt} \cite{balaprakashytopt} is an evolution of the work in \cite{sreenivasan2019framework}, that iterates over a set of user-defined configurations and their possible values to arrive at a tuned configuration.
These approaches are often domain or application specific.
Although often faster than search-based alternatives, these do need multiple code executions as evidenced by our experiments with {\tt ytopt} \cite{balaprakashytopt}.

Deep learning provides another alternative to the aforementioned techniques.
A suitable code representation technique is essential for such deep learning based code modeling. 
To this end, several code representations have been proposed \cite{venkatakeerthy2020ir2vec, cummins2021programl, cummins2020deep, raychev2015predicting, allamanis2017learning, alon2018general, brauckmann2020compiler, dam2018deep, ben2018neural}, which have been used to good effect for several optimization tasks such as heterogeneous device mapping, thread coarsening factor, etc. to name a few.
PROGRAML \cite{cummins2021programl} and IR2Vec \cite{venkatakeerthy2020ir2vec} are two state-of-the-art such code representations, which have addressed the shortcomings of seminal works in code representation such as {\tt inst2vec} \cite{ben2018neural}.
However, as mentioned before, each of these representations suffer from some limitations.
Given the complexity of developing new code representation techniques, building on top of existing ones seems wise.
Unlike the papers mentioned before, this study considers two code representations as two separate modalities for improving performance over unimodal approaches.

We have modeled our modalities using heterogeneous GNNs and DAEs.
A few works such as \cite{cummins2021programl, tehranijamsaz2022learning, dutta2022pattern, dutta2023power, tehranijamsaz2023paragraph}, in the recent past have successfully used GNNs for code modeling tasks.
However, to the best of our knowledge, this is the first work that employs heterogeneous GNNs for such tasks.
Additionally, we believe no other work has previously used denoising autoencoders to model code vectors and adapted multimodal learning for code representation learning.
\section{Discussion}
\label{sec:discussion}
Through this work, we have presented the idea of using heterogeneous GNNs, denoising autoencoders, and multimodal deep learning for the purpose of DL-based code modeling.
We believe the next phase of innovation in performance optimization will come from deep learning approaches.
As evidenced by works such as \cite{cummins2017end, cummins2021programl, ben2018neural, venkatakeerthy2020ir2vec, tehranijamsaz2022learning}, deep learning has been successfully used for compiler and performance optimizations.
Our experimental results in this paper further reinforce that belief.
However, DL is not a "silver bullet" for all problems.
These approaches do come with overheads in model training.
To address this, we have consciously designed very shallow networks to speed up training and inference at source.
The heterogeneous GNN model consists of only two hidden layers, the DAE model consists of only three hidden layers, and the MLP layers have one hidden layer.
This led to very fast training times of $\sim4$ seconds per epoch on a 10-core Intel SkyLake CPU.
Addtionally, profiling is an overhead for the experiments in Section \ref{sec:openmp_tuning}.
However, performance counters were a more automated way of incorporating the impact of inputs on code execution than handcrafting features.
In real world scenarios, handcrafting features might not be feasible due to the need of expert intervention.
But to alleviate the overhead, we narrowed down the collected counters to only five to reduce profiling overhead and to have a small feature set for faster training.
As shown in Section \ref{sec:openmp_tuning}, these steps led to faster tuning than existing autotuners.
\section{Conclusion and Future Works}
\label{sec:conclusions}
The presented technique of utilizing varied code representations as different modalities is unique and promising for optimization tasks.
The multimodal code representation outperforms both unimodal code representations considered in this paper.
Our multimodal learner also performs well when faced with unknown code and inputs.
This technique has led to us setting state-of-the-art results in the task of {\tt OpenCL} device mapping, and to the development of a multimodal {\tt OpenMP} tuner, producing better results than existing auto-tuners.
We aim to incorporate transfer and reinforcement learning in future efforts for developing an online tuner with customizable search spaces and expand our work to GPUs and FPGAs.

\begin{acks}
This research was supported by the National Science Foundation under Grant number 2211982.
We would also like to thank the ResearchIT team \footnote{https://researchit.las.iastate.edu} at Iowa State University for their constant support. This work was also supported by the Ministerio de Ciencia e Innovaci\'{o}n MCIN AEI/10.13039/501100011033 under contract PID2020-113614RB-C21 and by the Catalan government under contract 2021 SGR 00574.
\end{acks}

\bibliographystyle{ACM-Reference-Format}

\bibliography{hpdc.bib} 



\end{document}